\documentclass[a4paper,11pt]{article}
\usepackage{amssymb,amsthm}
\usepackage[leqno]{amsmath}
\usepackage{setspace}

\usepackage{tikz}
\usetikzlibrary{patterns}
\usetikzlibrary{decorations.pathreplacing,angles,quotes}

\usepackage[authoryear]{natbib}

\newcommand{\R}{\hbox{$ \rm I\hskip -0.14em R$}}

\newtheorem{cor}{Corollary}
\newtheorem{thm}{Theorem}
\newtheorem{lemma}{Lemma}
\newtheorem{prop}{Proposition}
\newcommand{\vs}[1]{{\vspace {#1cm}}}

\onehalfspace
\begin{document}

\title{%A necessary (and almost sufficient) condition for completely mixed Nash equilibria in finite two player games
%An implication of Farkas' lemma on the existence of completely mixed best-responses
Farkas' Lemma and Complete Indifference
\thanks{The research leading to these results has received funding from the People Programme (Marie Curie Actions) of the European Union's Seventh Framework Programme (FP7/2007-2013) under REA grant agreement PCIG11-GA-2012-322253. Florian Herold gratefully acknowledges this support.}
}
\author{Florian Herold\thanks{Department of Economics and Social Sciences, University of Bamberg, Feldkirchenstr. 21, 96045 Bamberg,
Germany, florian.herold@uni-bamberg.de} ~and Christoph Kuzmics\thanks{University of Graz, christoph.kuzmics@uni-graz.at }}

\date{\today}

\maketitle

\vs{1}

\begin{abstract}
%For arbitrary finite two-player games we establish a necessary and sufficient condition in terms of the payoff matrix of player~1 such that there exists a mixed strategy of the opponent player~2 which makes player~1 indifferent between all his strategies. The derivation is an application of Farkas' lemma. Since this condition is also a necessary condition for playing a completely mixed strategy we provide as a corollary a necessary (and almost sufficient) condition for the existence of completely mixed Nash-equilibria.  

%We establish a necessary (and almost sufficient) condition for an arbitrary finite symmetric two player game to have a symmetric completely mixed Nash equilibrium. The condition is a straightforward application of Farkas' lemma. We demonstrate its usefulness by providing the class of all symmetric two player three strategy games that have a unique and completely mixed symmetric Nash equilibrium. 

In a finite two player game consider the matrix of one player's payoff difference between any two consecutive pure strategies. Define the half space induced by a column vector of this matrix as the set of vectors that form an obtuse angle with this column vector. We use Farkas' lemma to show that this player can be made indifferent between all pure strategies if and only if the union of all these half spaces covers the whole vector space. This result leads to a necessary (and almost sufficient) condition for a game to have a completely mixed Nash equilibrium. We demonstrate its usefulness by providing the class of all symmetric two player three strategy games that have a unique and completely mixed symmetric Nash equilibrium. 
\end{abstract}


JEL-Code: C72

%\newpage

%\section{Introduction}
%
%
%

\section{Definitions and Results}

Consider an arbitrary finite two player game in which one of the two players has an $n \times m$ payoff matrix $A$. Let $x = (x_1, \ldots, x_m)^T \in \R^m$ with $x_i \ge 0$ for all $i \in \{1,2,...,m\}$ and $\sum_{i=1}^{m}x_i = 1$ denote a mixed strategy of this player's opponent. The player is only prepared to randomize over all their pure strategies if they all earn the same expected payoff. More precisely, $x$ has to be such that
\begin{equation}
Ax = (c,c,\ldots,c)^T ,
\end{equation}
for some constant $c \in \R$.

For any $n \times m$ matrix $A$, let $D=D(A)$ denote the $A$-induced payoff difference matrix given by the $(n-1) \times m$ matrix obtained from $A$ as follows. The $k$-th row of $D$ is the difference between rows $k$ and $k+1$ of matrix $A$, for $k=1,2,...,n-1$. Further, denote by $\bar{D}=\bar{D}(A)$ the $n \times m$ matrix that coincides with $D$ for the first $n-1$ rows and has the unit vector (vector of all ones) in row $n$ and define $b=(0,0,...,0,1)^T~\in~\R^n$.

Generally, for any vector $x \in  \R^k$ and any real number $a \in \R$ we write $x \ge a$ if $x_i \ge a$ for all $1 \leq i \leq k$, and we write $x \le a$ if $x_i \le a$ for all $1 \leq i \leq k$. We write $x>a$ if $x_i > a$ for all $1 \leq i \leq k$ and we write $x<a$ if $x_i < a$ for all $1 \leq i \leq k$.

\begin{lemma} \label{lemma_charcterize_completely mixed_eq} [Equal Payoff Condition] An opponent mixed strategy $x \in \R^m$ makes a player with payoff matrix $A$ indifferent between all pure strategies if and only if $\bar{D}x=b$ and $x \ge 0$.
\end{lemma}
\noindent The proof follows directly from the definitions of $\bar{D}$ and $b$. 

For $n-1 \times m$ matrix $D$ let $\mbox{col}(D)$ denote the set of column vectors of $D$. For any vector $d \in \R^{n-1}$ let $\mbox{HS}(d)$ denote the \emph{half space} induced by $d$, given by the set of all vectors $v \in \R^{n-1}$ such that $v^T d \le 0$. Furthermore, let $\mbox{HS}(A)=\bigcup_{d \in \text{col}(D)} \mbox{HS}(d)$ denote the union of all half spaces of columns of $D$. Note that $v \in \mbox{HS}(D)$ if and only if there is a $d \in \text{col}(D)$ such that $v^T d \le 0$, i.e. the angle between $v^T$ and $d$ is obtuse. 

\begin{thm} \label{thm_Cond_for_Cond_I}
Consider a finite two player normal form game in which one of the two players has an $n \times m$ payoff matrix $A$. This player can be made indifferent between all pure strategies if and only if $\mbox{HS}(D(A))=\R^{n-1}$, i.e.\ the union of half-spaces induced by the set of columns of the payoff difference matrix~$D$ covers the whole set $\R^{n-1}$.
\end{thm}

Note that we could also define the half space induced by any $d \in \text{col}(D)$ by the set of all vectors $v \in \R^{n-1}$ such that $v^T d \ge 0$ (the complement of the half space as defined above) and the Theorem could then be equivalently stated in terms of the unions of these half spaces.

\begin{cor} \label{cor_nec_cond_completelymixed_Nash_eq}
Consider a finite two player normal form game with $n_i$ pure strategies and payoff matrices $A_i$ for each player $i \in\{1,2\}$. Let $D_i = D(A_i)$ be the matrix, in which each row $k$ is given by the difference between row $k$ and $(k+1)$ in matrix $A_i$. Let $\mbox{HS}(d_i)$ denote the \emph{half space} induced by the column vector $d_i$ of matrix $D_i$, that is the set of all vectors $v \in \R^{n_{i}-1}$ such that $v^T d_i \le 0$. Let $\text{col}(D_i)$ denote the set of column vectors of $D_i$, and let $\mbox{HS}(D_i)=\bigcup_{d_i \in \text{col}(D_i)} \mbox{HS}(d_i)$ denote the union of all half spaces of columns of $D_i$. This game has a completely mixed Nash equilibrium only if $\mbox{HS}(D_i)=\R^{n_i-1}$ for both $i=1,2$, i.e.\ the union of half-spaces induced by the set of columns of the payoff difference matrix covers the whole set $\R^{n_i-1}$.
\end{cor}

Corollary~\ref{cor_nec_cond_completelymixed_Nash_eq} follows immediately from Lemma~\ref{lemma_charcterize_completely mixed_eq} and Theorem~\ref{thm_Cond_for_Cond_I}. 

%The result generalizes directly to general (possibly asymmetric) finite two player games. A general $n \times n$ bimatrix game with payoff matrices $A$ for player 1 and $B$ for player 2 has a completely mixed Nash equilibrium profile only if The condition in Corollary \ref{cor_nec_cond_completelymixed_Nash_eq} is met for both matrices $A$ and $B$.

\section{Proof of Theorem \ref{thm_Cond_for_Cond_I}}

The following lemma characterizes when there exists a solution~$x$ satisfying the Equal Payoff Condition:

\begin{lemma} \label{lemma_farkas_characterization_Cond_I}
There exists a vector $x \in \R^m$ with $x \geq 0$ such that $\bar{D}x = b$ if and only if there does not exist a vector $w \in \R^{n-1}$ such that $w^T D > 0$.
\end{lemma}

Note that the Lemma could also be stated by replacing the second condition by ``there does not exist a vector $w \in \R^{n-1}$ such that $w^T D < 0$.'' If there is a $w$ such that $w^T D < 0$ then there is $\tilde{w}=-w$ such that $\tilde{w}^T D > 0$. \medskip

\noindent {\bf Proof of Lemma~\ref{lemma_farkas_characterization_Cond_I}}: Farkas' lemma states that either there is an $x \in \R^m$ with $x \geq 0$ such that $\bar{D}x=b$ or there is a $v \in \R^n$ such $v^T \bar{D} \leq 0$ and $v^T b > 0$, but not both.\footnote{See \cite{farkas1902theorie} or e.g., \cite{Vohra05} for a textbook treatment of Farkas' lemma.} \medskip

``Only if'': By Farkas' lemma the existence of a solution to the Equal Payoff Condition implies that there is no $v=(v_1,v_2,...,v_n)^T \in \R^{n}$ with $v^T b > 0$ such that $v^T \bar{D} \le 0$. Given $b=(0,...,0,1)$, $v^T b = v_n$, and $v^T b > 0$ is satisfied if and only if $v_n > 0$. Let $w$ be the vector in $\R^{n-1}$ that consists of the first $n-1$ coordinates of $v$. Note that the condition $v^T \bar{D} \le 0$ is satisfied if and only if $w^T D \le - v_n$. 

Thus, the existence of a solution to the Equal Payoff Condition implies that there is no $(w_1,\ldots, w_{n-1},v_n)$ such that $w^T D \leq -v_n$ with $v_n >0$. This implies that there is no $w\in \R^{n-1}$ with $w^T D <0$. This, in turn, implies that there is no $w\in \R^{n-1}$ with $w^T D > 0$. If there were such a $w$ with $w^T D > 0$ then $\tilde{w}=-w$ satisfies $\tilde{w}^T D < 0$. \medskip

``If'': Suppose there is no vector $w \in \R^{n-1}$ such that $w^T D > 0$. Then there is no vector $w \in \R^{n-1}$ such that $w^T D < 0$. Then for all $v_n > 0$ there is no vector $w \in \R^{n-1}$ such that $w^T D \leq -v_n$. This implies that there is no vector $v=(v_1,\ldots, v_{n-1},v_n) \in \R^n$ such that $v^T \bar{D} \leq 0, v^T b >0$. Then Farkas' lemma implies the Equal Payoff Condition.
\hfill{QED} \bigskip

\noindent {\bf Proof of Theorem \ref{thm_Cond_for_Cond_I}}: 

``Only if:'' The existence of a solution~$x$ to the Equal Payoff Condition implies, by Lemma~\ref{lemma_farkas_characterization_Cond_I} that for every $w \in \R^{n-1}$ there must exist a vector $d \in \mbox{col}(D)$ such that $w^T d \leq 0$.
Thus, $w \in \mbox{HS}(D)$. This implies $\mbox{HS}(D)=\R^{n-1}$. \medskip

``If:'' Suppose $\mbox{HS}(D(A))=\R^{n-1}$. Then for any $w \in \R^{n-1}$ there is a $d \in \mbox{col}(D)$ such that $w^T d \leq 0$. Thus, there is no $w \in \R^{n-1}$ such that $w^T D > 0$. Then, by Lemma \ref{lemma_farkas_characterization_Cond_I}, there is a solution $x$ to the Equal Payoff Condition. \hfill{QED} \bigskip

\section{Examples}

\subsection{Two strategy games}

Consider an arbitrary symmetric two player two strategy game with payoff matrix

\[
A=\left(
\begin{array}{cc}
a & b \\
c & d
\end{array} \right),
\]
with 
\[
D(A)=\left(
\begin{array}{cc}
a-c & b-d \\
\end{array} \right).
\]
Then by Theorem~\ref{thm_Cond_for_Cond_I} such a game has a symmetric Nash equilibrium that satisfies the Equal Payoff Condition if and only if either $a-c > 0$ and $b-d <0$  or $a-c<0$ and $b-d>0$ (for $HS(D)=\R$). In the first case the game is a coordination game, in the second case a hawk-dove game, the only two classes (of four) symmetric two player two strategy games with a completely mixed Nash equilibrium. 

By the same argument, a general (possibly asymmetric) two player two strategy game has a symmetric Nash equilibrium only if the game is a coordination game, a hawk-dove game, or, and this is the only addition over the symmetric game, a game of the matching-pennies variety. %: one player has $a_i-c_i > 0$ and $b_i-d_i <0$  and the other $a_j-c_j<0$ and $b_j-d_j>0$.

\subsection{Three strategy games}

In this section we identify all (generic) symmetric two player three strategy games that have a unique symmetric equilibrium and, furthermore, this unique symmetric equilibrium is in completely mixed strategies.

Such a game can be identified by a general payoff matrix
\[
A=\left(
\begin{array}{ccc}
a_{11} & a_{12} & a_{13} \\
a_{21} & a_{22} & a_{23} \\
a_{31} & a_{32} & a_{33}
\end{array} \right),
\]
with all $a_{ji} \in \R$. Throughout we will make the (generic) assumption that $a_{ji} \ne a_{j'i}$ for all $j,j' \in \{1,2,3\}$ with $j \ne j'$. The best-response correspondence is unaffected if we subject payoffs to an affine transformation. This means we can, w.l.o.g., choose $\min_{j \in \{1,2,3\}} a_{ji} = 0$ for all $i \in \{1,2,3\}$ and $\max_{i,j \in \{1,2,3\}} a_{ji} = 1$. The existence of a symmetric completely mixed Nash equilibrium is also unaffected by choosing $\max_{j \in \{1,2,3\}} a_{ji} = 1$ for all $i \in \{1,2,3\}$. This is so because an opponent-strategy specific scalar multiplication of the payoff column in matrix $A$ implies a shortening or lengthening of the column vectors of the payoff-difference matrix $D$ without affecting their direction. Such a transformation will, therefore, not affect the half-spaces induced by the respective column vectors. Thus, without loss of generality, every column $i \in \{1,2,3\}$ has one $0$ payoff entry, one $1$ payoff entry, and one payoff entry $a_i$ with $0<a_i<1$.

%We shall consider two games equivalent if there is a permutation of strategies of one game that leads to a payoff matrix that is the same of the other game.

\begin{prop} \label{prop:allthree}
A symmetric two player three strategy game has 1) a unique symmetric equilibrium and 2) this equilibrium is in completely mixed strategies if and only if its payoff matrix $A$ can be transformed (using the transformations above) to one of the following six matrices (subject to strategy relabelling) with $a_i \in (0,1)$ for all $i \in \{1,2,3\}$:
\[\begin{array}{ccc}

A_1=\left(
\begin{array}{ccc}
0 & 1 & a_3 \\
a_1 & 0 & 1 \\
1 & a_2 & 0
\end{array} \right)

&

A_2=\left(
\begin{array}{ccc}
0 & 1 & 1 \\
a_1 & 0 & a_3 \\
1 & a_2 & 0
\end{array} \right)

&

A_3=\left(
\begin{array}{ccc}
a_1 & 1 & 0 \\
0 & a_2 & 1 \\
1 & 0 & a_3
\end{array} \right) \\

& \mbox{with } a_1 + a_3 > 1 & \\ \\

A_4=\left(
\begin{array}{ccc}
0 & 1 & 1 \\
1 & 0 & 0 \\
a_1 & a_2 & a_3
\end{array} \right)

&

A_5=\left(
\begin{array}{ccc}
0 & a_2 & 1 \\
1 & 0 & 0 \\
a_1 & 1 & a_3
\end{array} \right)

&

A_6=\left(
\begin{array}{ccc}
0 & 1 & 0 \\
a_1 & a_2 & 1 \\
1 & 0 & a_3
\end{array} \right) \\

\mbox{with } a_3 < 1-a_1 < a_2 & \mbox{with } a_1 + a_3 < 1 & \mbox{with } a_1 + a_2 < 1 
\end{array}
\]
\end{prop}

\begin{figure}[htb]
\begin{center}
\begin{tikzpicture}[scale=1]

\draw (4,5.5) node {$\mathcal{A}_1$};
\draw (7,5.5) node {$\mathcal{A}_2$};
\draw (2,7.5) node {$\mathcal{A}_4$};
\draw (2,3.5) node {$\mathcal{A}_5$};
\draw (9,7.5) node {$\mathcal{A}_3$};
\draw (9,3.5) node {$\mathcal{A}_6$};

\draw[-] (4.5,5.5) -- (6.5,5.5);
\draw[-] (2.5,7) -- (3.5,6);
\draw[-] (2.5,4) -- (3.5,5);
\draw[-] (2,4) -- (2,7);
\draw[-] (8.5,7) -- (7.5,6);
\draw[-] (8.5,4) -- (7.5,5);
\draw[-] (9,4) -- (9,7);

\end{tikzpicture}
\end{center}
\caption{\label{fig:adjacency} The adjacency relation between games classes $\mathcal{A}_1$ to $\mathcal{A}_6$.}
\end{figure}
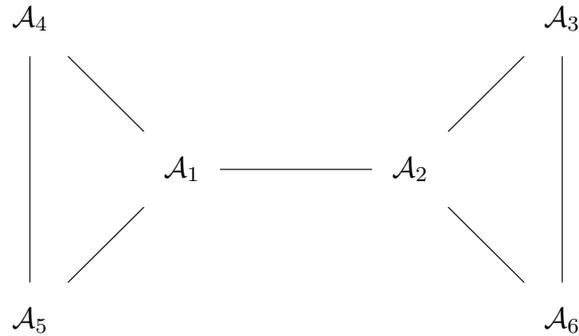

Note that each class of games $\mathcal{A}_i$ is a convex set. Moreover, no strictly convex combination of one matrix in one class $\mathcal{A}_i$ and another matrix in another class $\mathcal{A}_j$ ($j \ne i$) is in any of the six classes. One can define an adjacency relation $\sim$ with $\mathcal{A}_i \sim \mathcal{A}_j$ if there is a matrix that is in the intersection of the closure of both sets. Figure \ref{fig:adjacency} shows the resulting network of classes based on the adjacency relation. One implication is that the set of all symmetric 3 by 3 games with a unique Nash equilibrium and that is in completely mixed strategies is a connected component within the set of all 3 by 3 games that consists of minimally six convex subsets. 

Furthermore, note that all these games have a unique (symmetric) strategically stable set (in the sense of \cite{Kohlberg86}), this set is a singleton and includes exactly the unique Nash equilibrium (in completely mixed strategies). Also any symmetric 3 by 3 game with a completely mixed ESS must be in one of these six classes (by a result in \cite{Weibull95} that a game with a completely mixed ESS cannot have another symmetric equilibrium). 

\subsection{Proof of Proposition \ref{prop:allthree}}

Note first that diagonal matrix entries $a_{ii} \ne 1$ for all $i \in \{1,2,3\}$. Otherwise strategy $i$ is a symmetric pure strategy Nash equilibrium. Thus each $a_{ii}$ is either equal to $0$ or to $a_i \in (0,1)$. We are now going through the four basic cases of feasible matrices. \medskip

\paragraph{Case 0} The diagonal elements $a_{ii}=0$ for all $i \in \{1,2,3\}$. Without loss of generality (considering strategy relabelling) we can assume that the matrix is of the following structure.

\[A=\left(
\begin{array}{ccc}
0 &  &  \\
a_1 & 0 &  \\
1 &  & 0
\end{array} \right)\] \medskip

\noindent Case 0.1: We fill the second and third columns as follows, which gives us $A_1$:

\[A_1=\left(
\begin{array}{ccc}
0 & 1 & a_3 \\
a_1 & 0 & 1 \\
1 & a_2 & 0
\end{array} \right)\]
with
\[
D(A_1)=\left(
\begin{array}{ccc}
-a_1 & 1 & a_3-1 \\
a_1-1 & -a_2 & 1 \\
\end{array} \right).
\]

\begin{figure}[htb]
\begin{center}
\begin{tikzpicture}[scale=1]

\draw[->] (5.5,0) -- (5.5,11);
\draw[->] (0,5.5) -- (11,5.5);

\draw[-] (0.5,5.5) -- (5.5,0.5);

\draw[->,red,thick] (5.5,5.5) -- (3,3); 
\draw (3.5,4) node {$d_1$};
\draw[->] (4.6,4.4) -- (5.4,3.6); 
\draw[->] (4.4,4.6) -- (3.6,5.4); 

\draw[-,red] (5.5,0.5) -- (5.3,0.3);
\draw[-,red] (5.5,1) -- (5.3,0.8);
\draw[-,red] (5.5,1.5) -- (5.3,1.3);
\draw[-,red] (5.5,2) -- (5.3,1.8);
\draw[-,red] (5.5,2.5) -- (5.3,2.3);
\draw[-,red] (5.5,3) -- (5.3,2.8);
\draw[-,red] (5.5,3.5) -- (5.3,3.3);
\draw[-,red] (5.5,4) -- (5.3,3.8);
\draw[-,red] (5.5,4.5) -- (5.3,4.3);
\draw[-,red] (5.5,5) -- (5.3,4.8);

\draw[-,red] (5.5,5.5) -- (5.3,5.3);

\draw[-,red] (5,5.5) -- (4.8,5.3);
\draw[-,red] (4.5,5.5) -- (4.3,5.3);
\draw[-,red] (4,5.5) -- (3.8,5.3);
\draw[-,red] (3.5,5.5) -- (3.3,5.3);
\draw[-,red] (3,5.5) -- (2.8,5.3);
\draw[-,red] (2.5,5.5) -- (2.3,5.3);
\draw[-,red] (2,5.5) -- (1.8,5.3);
\draw[-,red] (1.5,5.5) -- (1.3,5.3);
\draw[-,red] (1,5.5) -- (0.8,5.3);
\draw[-,red] (0.5,5.5) -- (0.3,5.3);

\draw[-] (10.5,5.5) -- (10.5,0.5);
\draw[-] (5.5,5.5) -- (10.5,0.5);
\draw[->,blue,thick] (5.5,5.5) -- (10.5,3); 
\draw (9,4.1) node {$d_2$};
\draw[->] (8,4.35) -- (8,5.4); 
\draw[->] (8,4.15) -- (8,3.1); 

\draw[-] (5.5,5.5) -- (10.5,10.5);

\draw[-,blue] (5.5,5.5) -- (5.7,5.3);

\draw[-,blue] (5.5,5) -- (5.7,4.8);
\draw[-,blue] (5.5,4.5) -- (5.7,4.3);
\draw[-,blue] (5.5,4) -- (5.7,3.8);
\draw[-,blue] (5.5,3.5) -- (5.7,3.3);
\draw[-,blue] (5.5,3) -- (5.7,2.8);
\draw[-,blue] (5.5,2.5) -- (5.7,2.3);
\draw[-,blue] (5.5,2) -- (5.7,1.8);
\draw[-,blue] (5.5,1.5) -- (5.7,1.3);
\draw[-,blue] (5.5,1) -- (5.7,0.8);
\draw[-,blue] (5.5,0.5) -- (5.7,0.3);

\draw[-,blue] (6,6) -- (6.2,5.8);
\draw[-,blue] (6.5,6.5) -- (6.7,6.3);
\draw[-,blue] (7,7) -- (7.2,6.8);
\draw[-,blue] (7.5,7.5) -- (7.7,7.3);
\draw[-,blue] (8,8) -- (8.2,7.8);
\draw[-,blue] (8.5,8.5) -- (8.7,8.3);
\draw[-,blue] (9,9) -- (9.2,8.8);
\draw[-,blue] (9.5,9.5) -- (9.7,9.3);
\draw[-,blue] (10,10) -- (10.2,9.8);
\draw[-,blue] (10.5,10.5) -- (10.7,10.3);

\draw[-] (0.5,10.5) -- (5.5,10.5);
\draw[-] (5.5,5.5) -- (0.5,10.5);
\draw[->,violet,thick] (5.5,5.5) -- (3,10.5); 
\draw (3.4,8.9) node {$d_3$};
\draw[->] (4.35,8) -- (5.4,8); 
\draw[->] (4.15,8) -- (3.1,8); 

\draw[-,violet] (5.5,5.5) -- (5.3,5.7);

\draw[-,violet] (5,5.5) -- (4.8,5.7);
\draw[-,violet] (4.5,5.5) -- (4.3,5.7);
\draw[-,violet] (4,5.5) -- (3.8,5.7);
\draw[-,violet] (3.5,5.5) -- (3.3,5.7);
\draw[-,violet] (3,5.5) -- (2.8,5.7);
\draw[-,violet] (2.5,5.5) -- (2.3,5.7);
\draw[-,violet] (2,5.5) -- (1.8,5.7);
\draw[-,violet] (1.5,5.5) -- (1.3,5.7);
\draw[-,violet] (1,5.5) -- (0.8,5.7);
\draw[-,violet] (0.5,5.5) -- (0.3,5.7);

\draw[-,violet] (6,6) -- (5.8,6.2);
\draw[-,violet] (6.5,6.5) -- (6.3,6.7);
\draw[-,violet] (7,7) -- (6.8,7.2);
\draw[-,violet] (7.5,7.5) -- (7.3,7.7);
\draw[-,violet] (8,8) -- (7.8,8.2);
\draw[-,violet] (8.5,8.5) -- (8.3,8.7);
\draw[-,violet] (9,9) -- (8.8,9.2);
\draw[-,violet] (9.5,9.5) -- (9.3,9.7);
\draw[-,violet] (10,10) -- (9.8,10.2);
\draw[-,violet] (10.5,10.5) -- (10.3,10.7);

%\draw (5.5,10.5) node[left] {$1$};
%\draw (10.5,5.5) node[below] {$1$};
%\draw (5.5,0.5) node[left] {$-1$};
%\draw (0.5,5.5) node[below] {$-1$};

\end{tikzpicture}
\end{center}
\caption{\label{fig:classA1} Class $A_1$ with $
D(A_1)=\left(
\begin{array}{ccc}
-a_1 & 1 & a_3-1 \\
a_1-1 & -a_2 & 1 \\
\end{array} \right).
$}
\end{figure}
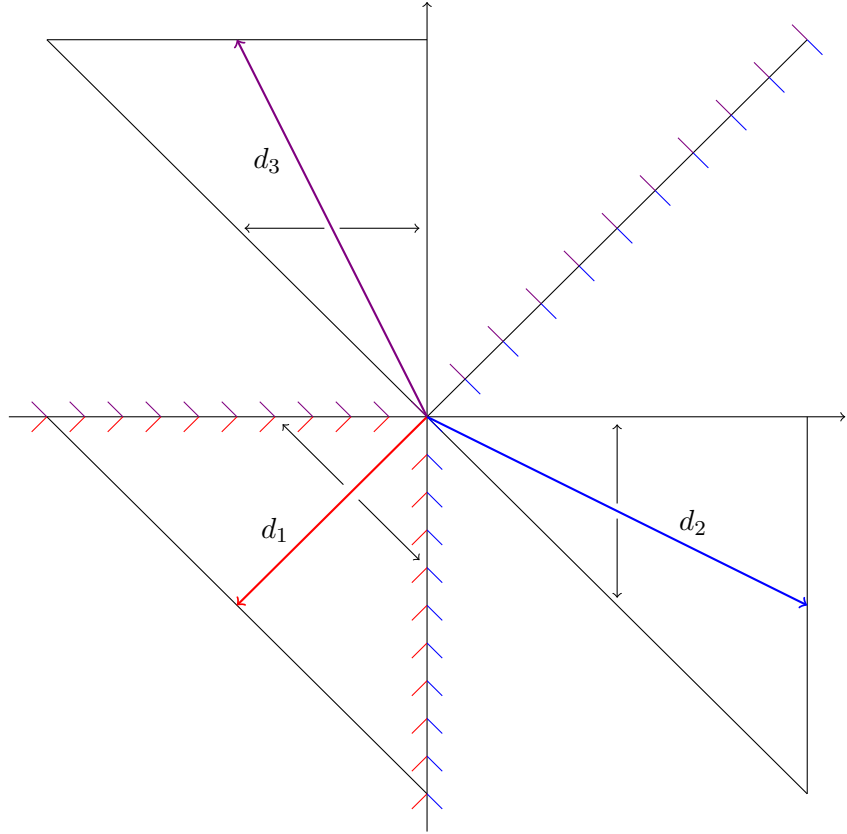

Figure \ref{fig:classA1} provides a geometric representation of $D(A_1)$: vector $d_i$ for $i \in \{1,2,3\}$ is the $i$-th column of matrix $D(A)$. The little side-ways arrows mark the feasible range for the three vectors, respectively, as the payoff parameters $a_i$ vary between $0$ and $1$. The half space generated by vector $d_1$ covers at least the area marked in red, while for $d_2$ it is the area marked in blue and for $d_3$ the area marked in violet. Clearly, the three half-spaces together cover all of $\R^2$ for any values $a_i \in (0,1)$ for $i \in \{1,2,3\}$. This proves, by Theorem \ref{thm_Cond_for_Cond_I}, that this game has a completely mixed symmetric equilibrium. To see that it is the only symmetric equilibrium, note that the game has no symmetric pure strategy equilibria and for any mixed strategy that uses only two pure strategies, the left out pure strategy then strictly dominates one of the two strategies. \medskip

\noindent Case 0.2: We fill the second and third columns as follows, which gives us $A_2$:

\[A_2=\left(
\begin{array}{ccc}
0 & 1 & 1 \\
a_1 & 0 & a_3 \\
1 & a_2 & 0
\end{array} \right)\]
with
\[
D(A_2)=\left(
\begin{array}{ccc}
-a_1 & 1 & 1-a_3 \\
a_1-1 & -a_2 & a_3 \\
\end{array} \right).
\]

\begin{figure}[htb]
\begin{center}
\begin{tikzpicture}[scale=1]

\draw[->] (5.5,0) -- (5.5,11);
\draw[->] (0,5.5) -- (11,5.5);

\draw[-] (0.5,5.5) -- (5.5,0.5);

\draw[->,red,thick] (5.5,5.5) -- (3,3); 
\draw (3.5,3.9) node {$d_1$};
\draw[->] (4.6,4.4) -- (5.4,3.6); 
\draw[->] (4.4,4.6) -- (3.6,5.4); 

\draw[-,red] (5.5,0.5) -- (5.3,0.3);
\draw[-,red] (5.5,1) -- (5.3,0.8);
\draw[-,red] (5.5,1.5) -- (5.3,1.3);
\draw[-,red] (5.5,2) -- (5.3,1.8);
\draw[-,red] (5.5,2.5) -- (5.3,2.3);
\draw[-,red] (5.5,3) -- (5.3,2.8);
\draw[-,red] (5.5,3.5) -- (5.3,3.3);
\draw[-,red] (5.5,4) -- (5.3,3.8);
\draw[-,red] (5.5,4.5) -- (5.3,4.3);
\draw[-,red] (5.5,5) -- (5.3,4.8);

\draw[-,red] (5.5,5.5) -- (5.3,5.3);

\draw[-,red] (5,5.5) -- (4.8,5.3);
\draw[-,red] (4.5,5.5) -- (4.3,5.3);
\draw[-,red] (4,5.5) -- (3.8,5.3);
\draw[-,red] (3.5,5.5) -- (3.3,5.3);
\draw[-,red] (3,5.5) -- (2.8,5.3);
\draw[-,red] (2.5,5.5) -- (2.3,5.3);
\draw[-,red] (2,5.5) -- (1.8,5.3);
\draw[-,red] (1.5,5.5) -- (1.3,5.3);
\draw[-,red] (1,5.5) -- (0.8,5.3);
\draw[-,red] (0.5,5.5) -- (0.3,5.3);

\draw[-] (10.5,5.5) -- (10.5,0.5);
\draw[-] (5.5,5.5) -- (10.5,0.5);
\draw[->,blue,thick] (5.5,5.5) -- (10.5,3); 
\draw (9,4.1) node {$d_2$};
\draw[->] (8,4.35) -- (8,5.4); 
\draw[->] (8,4.15) -- (8,3.1); 

\draw[-] (5.5,5.5) -- (10.5,10.5);

\draw[-,blue] (5.5,5.5) -- (5.7,5.3);

\draw[-,blue] (5.5,5) -- (5.7,4.8);
\draw[-,blue] (5.5,4.5) -- (5.7,4.3);
\draw[-,blue] (5.5,4) -- (5.7,3.8);
\draw[-,blue] (5.5,3.5) -- (5.7,3.3);
\draw[-,blue] (5.5,3) -- (5.7,2.8);
\draw[-,blue] (5.5,2.5) -- (5.7,2.3);
\draw[-,blue] (5.5,2) -- (5.7,1.8);
\draw[-,blue] (5.5,1.5) -- (5.7,1.3);
\draw[-,blue] (5.5,1) -- (5.7,0.8);
\draw[-,blue] (5.5,0.5) -- (5.7,0.3);

\draw[-,blue] (6,6) -- (6.2,5.8);
\draw[-,blue] (6.5,6.5) -- (6.7,6.3);
\draw[-,blue] (7,7) -- (7.2,6.8);
\draw[-,blue] (7.5,7.5) -- (7.7,7.3);
\draw[-,blue] (8,8) -- (8.2,7.8);
\draw[-,blue] (8.5,8.5) -- (8.7,8.3);
\draw[-,blue] (9,9) -- (9.2,8.8);
\draw[-,blue] (9.5,9.5) -- (9.7,9.3);
\draw[-,blue] (10,10) -- (10.2,9.8);
\draw[-,blue] (10.5,10.5) -- (10.7,10.3);

\draw[-] (5.5,10.5) -- (10.5,5.5);
\draw[->,violet,thick] (5.5,5.5) -- (8,8); 
\draw (7.5,6.9) node {$d_3$};
\draw[->] (6.6,6.4) -- (7.4,5.6); 
\draw[->] (6.4,6.6) -- (5.6,7.4); 

\draw[-,violet] (5.5,6) -- (5.7,6.2);
\draw[-,violet] (5.5,6.5) -- (5.7,6.7);
\draw[-,violet] (5.5,7) -- (5.7,7.2);
\draw[-,violet] (5.5,7.5) -- (5.7,7.7);
\draw[-,violet] (5.5,8) -- (5.7,8.2);
\draw[-,violet] (5.5,8.5) -- (5.7,8.7);
\draw[-,violet] (5.5,9) -- (5.7,9.2);
\draw[-,violet] (5.5,9.5) -- (5.7,9.7);
\draw[-,violet] (5.5,10) -- (5.7,10.2);
\draw[-,violet] (5.5,10.5) -- (5.7,10.7);

\draw[-,violet] (5.5,5.5) -- (5.7,5.7);

\draw[-,violet] (6,5.5) -- (6.2,5.7);
\draw[-,violet] (6.5,5.5) -- (6.7,5.7);
\draw[-,violet] (7,5.5) -- (7.2,5.7);
\draw[-,violet] (7.5,5.5) -- (7.7,5.7);
\draw[-,violet] (8,5.5) -- (8.2,5.7);
\draw[-,violet] (8.5,5.5) -- (8.7,5.7);
\draw[-,violet] (9,5.5) -- (9.2,5.7);
\draw[-,violet] (9.5,5.5) -- (9.7,5.7);
\draw[-,violet] (10,5.5) -- (10.2,5.7);
\draw[-,violet] (10.5,5.5) -- (10.7,5.7);

%\draw[->,lightgray,thick] (5.5,5.5) -- (7.5,8.5); 
\draw[->,lightgray,thick] (5.5,5.5) -- (9,7); 
\draw[-] (9,7) -- (9,5.5);
\draw (9,6.25) node[right]{$a_3$};

\draw[->,lightgray,thick] (5.5,5.5) -- (4,2); 
%\draw[->,lightgray,thick] (5.5,5.5) -- (2.5,3.5); 
\draw[-] (4,2) -- (5.5,2);
\draw (4.75,2) node[below]{$a_1$};

%\draw (5.5,10.5) node[left] {$1$};
%\draw (10.5,5.5) node[below] {$1$};
%\draw (5.5,0.5) node[left] {$-1$};
%\draw (0.5,5.5) node[below] {$-1$};

\end{tikzpicture}
\end{center}
\caption{\label{fig:classA2} Class $A_2$ with $
D(A_2)=\left(
\begin{array}{ccc}
-a_1 & 1 & 1-a_3 \\
a_1-1 & -a_2 & a_3 \\
\end{array} \right).
$}
\end{figure}
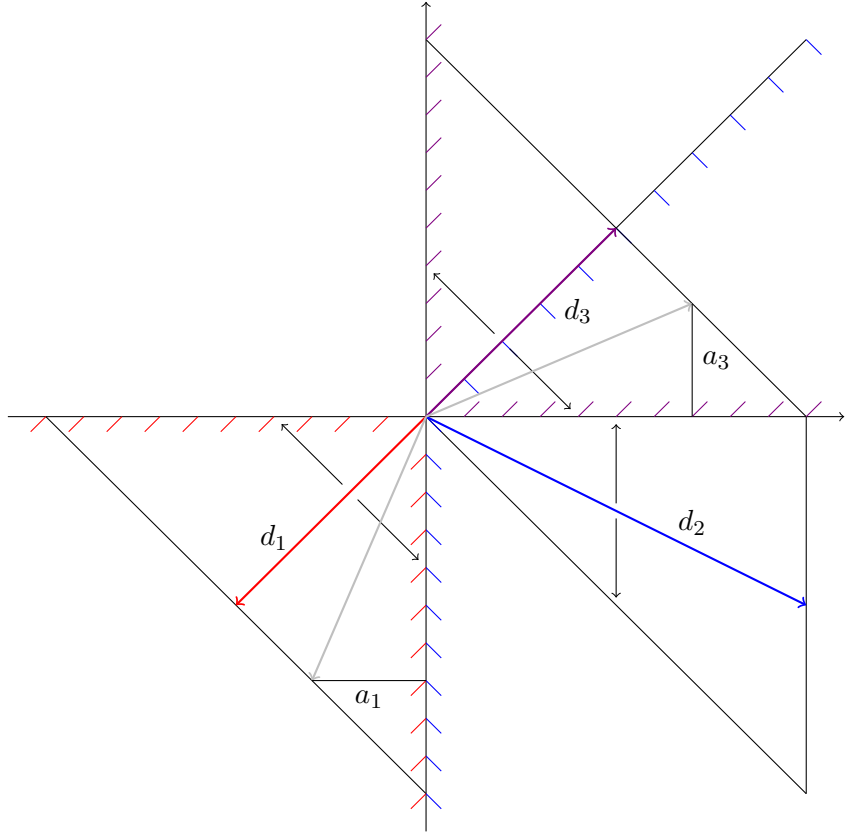

Figure \ref{fig:classA2} provides a geometric representation of $D(A_2)$, analogous to the previous case. The union of the three half-spaces $\mbox{HS}(d_i)$, for $i \in \{1,2,3\}$, do not necessarily cover the whole of $\R^2$. For instance if we have the two light gray versions of vectors $d_1$ and $d_3$ (with $a_1+a_3<1$) the union of all half-spaces leaves a gap in the upper left quadrant. In fact, it is easy to see that we get $\mbox{HS}(A)=\R^2$ if and only if $a_1+a_3>1$.\footnote{Note that if $a_1+a_3 <1$ strategy $2$ is strictly dominated by some appropriate mixture of strategies $1$ and $3$. The graphical argument, based on Corollary \ref{cor_nec_cond_completelymixed_Nash_eq} demonstrates that this is the only parameter configuration that we have to rule out in this case.} 

To see that the completely mixed symmetric equilibrium is the unique symmetric equilibrium we need to consider all strategies that use exactly two pure strategies. Suppose first that strategy $3$ is removed. But then strategy $3$ dominates strategy $2$, and, thus, there cannot be an equilibrium using only strategies $1$ and $2$. Suppose secondly that strategy strategy $2$ is removed, then the only candidate for a mixed equilibrium is one in which we attach equal probability of one half on both $1$ and $3$. In this case, however, as $a_1+a_3>1$, strategy $3$ is a better response. Suppose thirdly and finally that strategy $1$ is removed. Then strategy $1$ dominates both strategies $2$ and $3$. \medskip

\noindent Case 0.3: We fill the second and third columns as follows, which gives us 
\[\left(
\begin{array}{ccc}
0 & a_2 & a_3 \\
a_1 & 0 & 1 \\
1 & 1 & 0
\end{array} \right).\]
Relabelling strategies $3 \to 1 \to 2 \to 3$, we get
\[\left(
\begin{array}{ccc}
0 & 1 & 1 \\
a_3 & 0 & a_2 \\
1 & a_1 & 0
\end{array} \right),\]
which has the same structure as $A_2$. \medskip

\noindent Case 0.4: We fill the second and third columns as follows, which gives us 
\[\left(
\begin{array}{ccc}
0 & a_2 & 1 \\
a_1 & 0 & a_3 \\
1 & 1 & 0
\end{array} \right),\]
which after relabelling $3 \to 1 \to 3$ and $2 \to 2$ leads to
\[\left(
\begin{array}{ccc}
0 & 1 & 1 \\
a_3 & 0 & a_1 \\
1 & a_2 & 0
\end{array} \right),\]
which, again, has the same structure as $A_2$. \medskip

\paragraph{Case 1} One diagonal payoff matrix entry $a_{ii}=a_i$ and the other two are equal to zero. W.l.o.g., say $a_{33}=a_3$. Without loss of generality (considering strategy relabelling) we can assume that the matrix is of the following structure.

\[A=\left(
\begin{array}{ccc}
0 &  & 1 \\
 & 0 & 0 \\
 &  & a_3
\end{array} \right)\] \medskip

\noindent Case 1.1: We fill the first and second columns as follows, which gives us 
\[\left(
\begin{array}{ccc}
0 & a_2  & 1 \\
a_1 & 0 & 0 \\
1 & 1 & a_3
\end{array} \right).\]
In this case strategy $3$ strictly dominates strategy $2$ and the game cannot have a completely mixed equilibrium. This could also be seen by noting that all three column vectors of $D(A)$, in this case, have a negative second coordinate, and thus, the union of their half-spaces do not cover all of $\R^2$.\medskip

\noindent Case 1.2: We fill the first and second columns as follows, which gives us 
\[\left(
\begin{array}{ccc}
0 & 1  & 1 \\
a_1 & 0 & 0 \\
1 & a_2 & a_3
\end{array} \right).\]
As in the previous case, strategy $3$ strictly dominates strategy $2$ (and the union of half-spaces do not cover $\R^2$) and the game cannot have a completely mixed equilibrium.\medskip

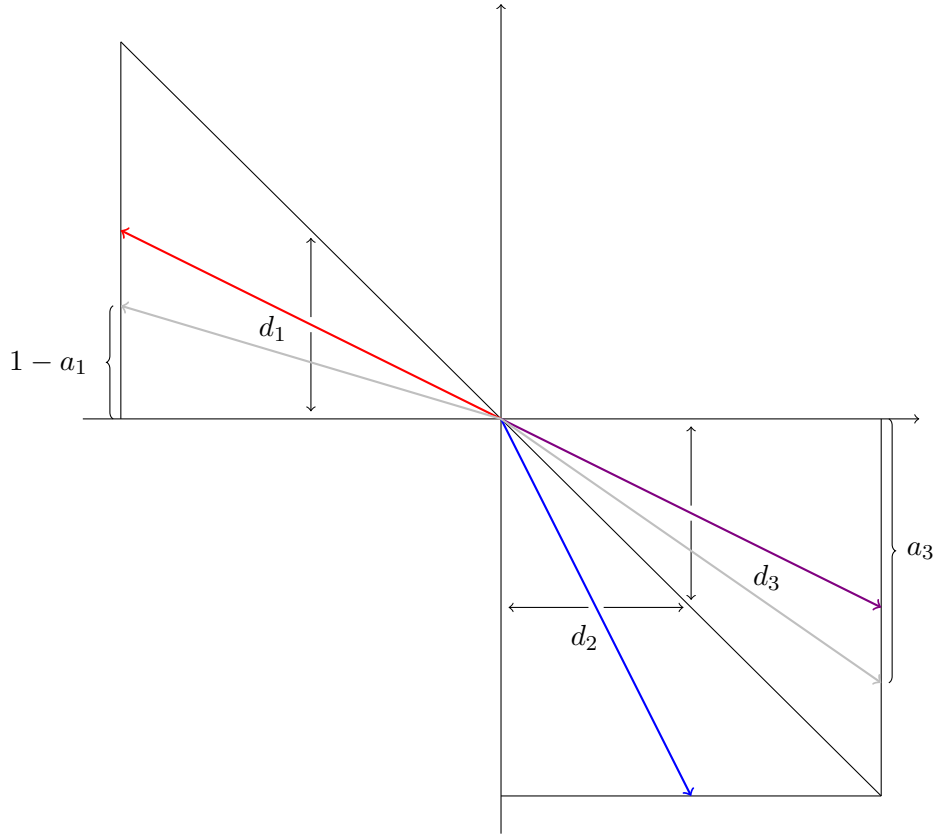
\begin{figure}[htb]
\begin{center}
\begin{tikzpicture}[scale=1]

\draw[->] (5.5,0) -- (5.5,11);
\draw[->] (0,5.5) -- (11,5.5);

\draw[-] (5.5,5.5) -- (0.5,10.5);
\draw[-] (0.5,5.5) -- (0.5,10.5);

\draw[->,red,thick] (5.5,5.5) -- (0.5,8); 
\draw (2.5,6.7) node {$d_1$};
\draw[->] (3,6.65) -- (3,5.6); 
\draw[->] (3,6.85) -- (3,7.9);

\draw[-] (10.5,5.5) -- (10.5,0.5);
\draw[-] (5.5,5.5) -- (10.5,0.5);
\draw[->,violet,thick] (5.5,5.5) -- (10.5,3); 
\draw (9,3.4) node {$d_3$};
\draw[->] (8,4.35) -- (8,5.4); 
\draw[->] (8,4.15) -- (8,3.1); 

%\draw[-] (5.5,5.5) -- (10.5,10.5);

\draw[-] (5.5,0.5) -- (10.5,0.5);
\draw[->,blue,thick] (5.5,5.5) -- (8,0.5); 
\draw (6.6,2.6) node {$d_2$};
\draw[->] (6.65,3) -- (5.6,3); 
\draw[->] (6.85,3) -- (7.9,3); 

\draw[->,lightgray,thick] (5.5,5.5) -- (0.5,7); 
\draw[->,lightgray,thick] (5.5,5.5) -- (10.5,2); 

\draw[decoration={brace,mirror},decorate] (10.6,2) -- (10.6,5.5);
\draw (10.7,3.75) node[right]{$a_3$};

\draw[decoration={brace},decorate] (0.4,5.5) -- (0.4,7);
\draw (0.2,6.25) node[left]{$1-a_1$};

\end{tikzpicture}
\end{center}
\caption{\label{fig:classA5} Class $A_5$ with $
D(A_5)=\left(
\begin{array}{ccc}
-1 & a_2 & 1 \\
1-a_1 & -1 & -a_3 \\
\end{array} \right).
$}
\end{figure}

\noindent Case 1.3: We fill the first and second columns as follows, which gives us $A_5$:
\[A_5=\left(
\begin{array}{ccc}
0 & a_2 & 1 \\
1 & 0 & 0 \\
a_1 & 1 & a_3
\end{array} \right)\]
with
\[
D(A_5)=\left(
\begin{array}{ccc}
-1 & a_2 & 1 \\
1-a_1 & -1 & -a_3 \\
\end{array} \right).
\]
No pure strategy dominates another, yet not all such games have a completely mixed symmetric equilibrium. Figure \ref{fig:classA5} provides a geometric representation of $D(A_5)$. The union of the three half-spaces $\mbox{HS}(d_i)$, for $i \in \{1,2,3\}$, do not necessarily cover the whole of $\R^2$. For instance, if we have the two light gray versions of vectors $d_1$ and $d_3$ (with $a_1+a_3>1$) the union of all half-spaces leaves a gap in the upper right quadrant. In fact, it is easy to see that we get $\mbox{HS}(A)=\R^2$ if and only if $a_1+a_3<1$.

To see that the completely mixed symmetric equilibrium is the unique symmetric equilibrium, we need to consider all strategies that use exactly two pure strategies. Suppose first that strategy $1$ is removed. Then strategy $3$ dominates strategy $2$. Suppose secondly that strategy $2$ is removed. Then for the only candidate symmetric equilibrium using strategies $1$ and $3$ only, strategy $2$ is a better response when $a_1 + a_3 < 1$. Finally, suppose that strategy $3$ is removed. Then strategy $3$ dominates strategy $1$. \medskip

\noindent Case 1.4: We fill the first and second columns as follows, which gives us $A_4$:
\[A_4=\left(
\begin{array}{ccc}
0 & 1 & 1 \\
1 & 0 & 0 \\
a_1 & a_2 & a_3
\end{array} \right)\]
with
\[
D(A_4)=\left(
\begin{array}{ccc}
-1 & 1 & 1 \\
1-a_1 & -a_2 & -a_3 \\
\end{array} \right).
\]
Figure \ref{fig:classA5} provides a geometric representation of $D(A_4)$. In order for the union of the three half-spaces $\mbox{HS}(d_i)$, for $i \in \{1,2,3\}$, to cover the whole of $\R^2$, we clearly need that either $a_2 < 1-a_1 < a_3$ or $a_3 < 1-a_1 < a_2$ (we need $-d_1$ to be in the convex cone generated by $d_2$ and $d_3$). 

Is this completely mixed symmetric equilibrium unique? The game certainly does not have a symmetric equilibrium using only pure strategies $2$ and $3$. The only candidate for a symmetric equilibrium using only pure strategies $1$ and $2$ is too mix equally between them. This is not an equilibrium if strategy $3$ is a better reply than both strategies $1$ and $2$, which is the case if and only if $\frac12 a_1+ \frac12 a_2 > \frac12$ or, equivalently, $a_1+a_2 > 1$ or $a_2 > 1-a_1$, thus eliminating the case $a_3 < 1-a_1 < a_2$. Finally, the game has a symmetric equilibrium using only strategies $1$ and $3$ if and only if $a_3 < 1-a_1$, which provides no additional restriction. \medskip

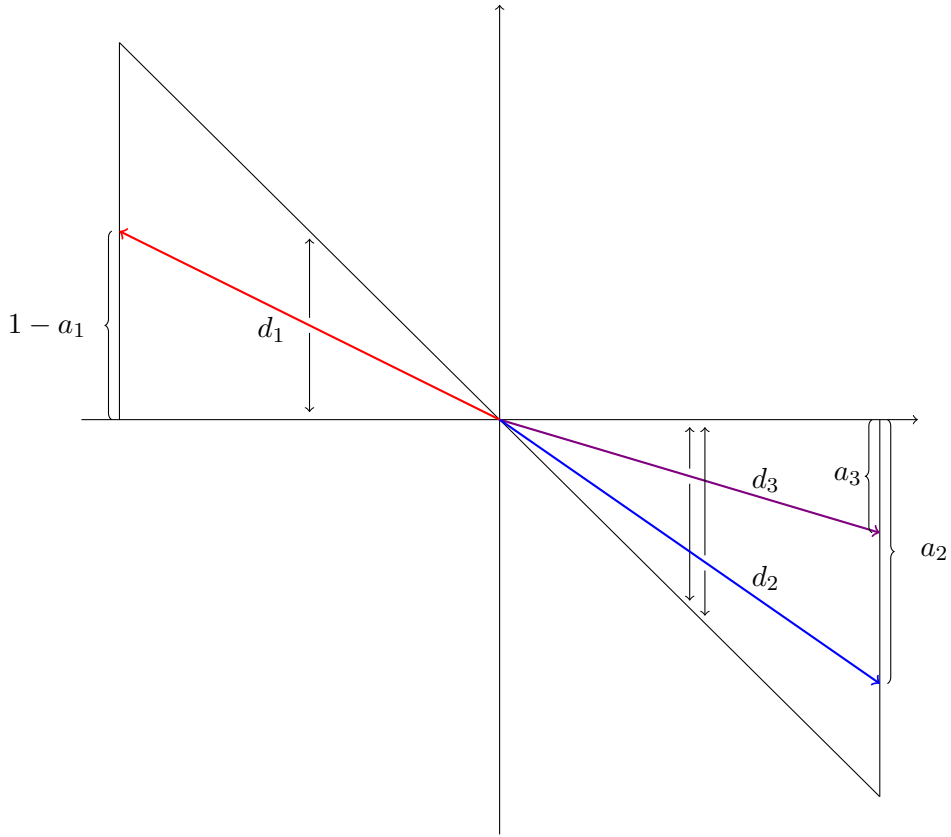
\begin{figure}[htb]
\begin{center}
\begin{tikzpicture}[scale=1]

\draw[->] (5.5,0) -- (5.5,11);
\draw[->] (0,5.5) -- (11,5.5);

\draw[-] (5.5,5.5) -- (0.5,10.5);
\draw[-] (0.5,5.5) -- (0.5,10.5);

\draw[->,red,thick] (5.5,5.5) -- (0.5,8); 
\draw (2.5,6.7) node {$d_1$};
\draw[->] (3,6.65) -- (3,5.6); 
\draw[->] (3,6.85) -- (3,7.9);

\draw[-] (10.5,5.5) -- (10.5,0.5);
\draw[-] (5.5,5.5) -- (10.5,0.5);
\draw[->,violet,thick] (5.5,5.5) -- (10.5,4); 
\draw (9,4.7) node {$d_3$};
\draw[->] (8,4.85) -- (8,5.4); 
\draw[->] (8,4.65) -- (8,3.1); 

%\draw[-] (5.5,5.5) -- (10.5,10.5);

\draw[->,blue,thick] (5.5,5.5) -- (10.5,2); 
\draw (9,3.4) node {$d_2$};
\draw[->] (8.2,3.7) -- (8.2,5.4); 
\draw[->] (8.2,3.5) -- (8.2,2.9);

\draw[decoration={brace,mirror},decorate] (10.6,2) -- (10.6,5.5);
\draw (10.9,3.75) node[right]{$a_2$};

\draw[decoration={brace},decorate] (10.4,4) -- (10.4,5.5);
\draw (10.4,4.75) node[left]{$a_3$};

\draw[decoration={brace},decorate] (0.4,5.5) -- (0.4,8);
\draw (0.2,6.75) node[left]{$1-a_1$};

\end{tikzpicture}
\end{center}
\caption{\label{fig:classA4} Class $A_4$ with $
D(A_4)=\left(
\begin{array}{ccc}
-1 & 1 & 1 \\
1-a_1 & -a_2 & -a_3 \\
\end{array} \right).
$}
\end{figure}

\paragraph{Case 2} One diagonal payoff matrix entry $a_{ii}=0$ and the other two $a_{jj}=a_j$ $(j \ne i)$. W.l.o.g., say $a_{11}=0$. Without loss of generality (considering strategy relabelling) we can assume that the matrix is of the following structure.

\[A=\left(
\begin{array}{ccc}
0 &  &  \\
a_1 & a_2 & \\
1 &  & a_3
\end{array} \right)\] \medskip

\noindent Case 2.1: We fill the second and third columns as follows, which gives us 
\[\left(
\begin{array}{ccc}
0 & 0 & 0 \\
a_1 & a_2 & 1 \\
1 & 1 & a_3
\end{array} \right)\]
In this case strategy $1$ is strictly dominated and there can be no completely mixed equilibrium. \medskip

\noindent Case 2.2: We fill the second and third columns as follows, which gives us 
\[\left(
\begin{array}{ccc}
0 & 0 & 1 \\
a_1 & a_2 & 0 \\
1 & 1 & a_3
\end{array} \right)\]
In this case strategy $2$ is strictly dominated and there can be no completely mixed equilibrium. \medskip

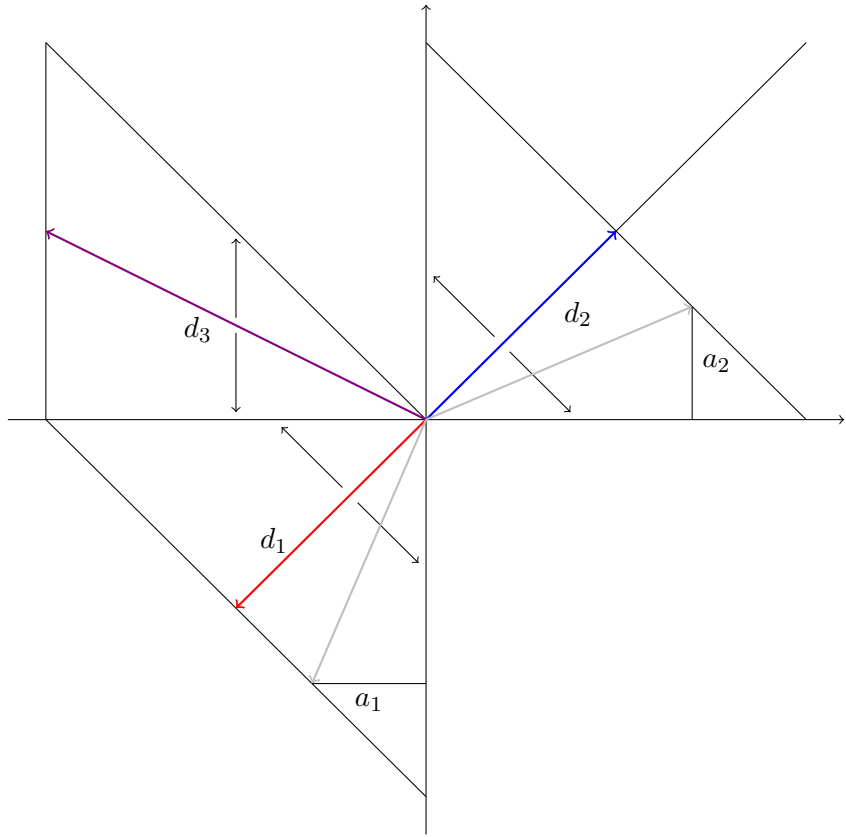
\begin{figure}[htb]
\begin{center}
\begin{tikzpicture}[scale=1]

\draw[->] (5.5,0) -- (5.5,11);
\draw[->] (0,5.5) -- (11,5.5);

\draw[-] (0.5,5.5) -- (5.5,0.5);

\draw[->,red,thick] (5.5,5.5) -- (3,3); 
\draw (3.5,3.9) node {$d_1$};
\draw[->] (4.6,4.4) -- (5.4,3.6); 
\draw[->] (4.4,4.6) -- (3.6,5.4);

\draw[-] (5.5,5.5) -- (0.5,10.5);
\draw[-] (0.5,5.5) -- (0.5,10.5);

\draw[->,violet,thick] (5.5,5.5) -- (0.5,8); 
\draw (2.5,6.7) node {$d_3$};
\draw[->] (3,6.65) -- (3,5.6); 
\draw[->] (3,6.85) -- (3,7.9);

\draw[-] (5.5,5.5) -- (10.5,10.5);

\draw[-] (5.5,10.5) -- (10.5,5.5);
\draw[->,blue,thick] (5.5,5.5) -- (8,8); 
\draw (7.5,6.9) node {$d_2$};
\draw[->] (6.6,6.4) -- (7.4,5.6); 
\draw[->] (6.4,6.6) -- (5.6,7.4); 

%\draw[->,lightgray,thick] (5.5,5.5) -- (7.5,8.5); 
\draw[->,lightgray,thick] (5.5,5.5) -- (9,7); 
\draw[-] (9,7) -- (9,5.5);
\draw (9,6.25) node[right]{$a_2$};

\draw[->,lightgray,thick] (5.5,5.5) -- (4,2); 
%\draw[->,lightgray,thick] (5.5,5.5) -- (2.5,3.5); 
\draw[-] (4,2) -- (5.5,2);
\draw (4.75,2) node[below]{$a_1$};

%\draw (5.5,10.5) node[left] {$1$};
%\draw (10.5,5.5) node[below] {$1$};
%\draw (5.5,0.5) node[left] {$-1$};
%\draw (0.5,5.5) node[below] {$-1$};

\end{tikzpicture}
\end{center}
\caption{\label{fig:classA6} Class $A_6$ with
$D(A_6)=\left(
\begin{array}{ccc}
-a_1 & 1-a_2 & -1 \\
a_1-1 & a_2 & 1-a_3 
\end{array} \right).
$}
\end{figure}

\noindent Case 2.3: We fill the second and third columns as follows, which gives us $A_6$:
\[A_6=\left(
\begin{array}{ccc}
0 & 1 & 0 \\
a_1 & a_2 & 1 \\
1 & 0 & a_3
\end{array} \right)\]
with
\[D(A_6)=\left(
\begin{array}{ccc}
-a_1 & 1-a_2 & -1 \\
a_1-1 & a_2 & 1-a_3 
\end{array} \right).\]
In this game there are no dominated strategies, yet this game does not have a completely mixed symmetric equilibrium for all parameter values. Figure \ref{fig:classA6} provides a geometric representation of $D(A_6)$. In order for the union of the three half-spaces $\mbox{HS}(d_i)$, for $i \in \{1,2,3\}$, to cover the whole of $\R^2$, we clearly need that $a_1 + a_2 < 1$. 

To see that the completely mixed symmetric equilibrium is the unique symmetric equilibrium, we need to consider all strategies that use exactly two pure strategies. Suppose first that strategy $1$ is removed. Then strategy $2$ dominates strategy $3$. Suppose secondly that strategy $2$ is removed. Then strategy $3$ dominates strategy $1$. Suppose, finally, that strategy $3$ is removed. Then, analogously to the case of game $A_4$, the only candidate equilibrium using only strategies $1$ and $2$ has strategy $3$ as a better response when $a_1+a_2 < 1$.\medskip

\noindent Case 2.4: We fill the second and third columns as follows, which gives us
\[\left(
\begin{array}{ccc}
0 & 1 & 1 \\
a_1 & a_2 & 0 \\
1 & 0 & a_3
\end{array} \right)\]
This has a non-completely mixed Nash equilibrium using only pure strategies $1$ and $3$. \medskip

\begin{figure}[h]
\begin{center}
\begin{tikzpicture}[scale=1]

\draw[->] (5.5,0) -- (5.5,11);
\draw[->] (0,5.5) -- (11,5.5);

\draw[-] (5.5,5.5) -- (0.5,10.5);
\draw[-] (0.5,5.5) -- (0.5,10.5);

\draw[->,violet,thick] (5.5,5.5) -- (0.5,8); 
\draw (2.5,6.7) node {$d_1$};
\draw[->] (3,6.65) -- (3,5.6); 
\draw[->] (3,6.85) -- (3,7.9);

\draw[-] (5.5,5.5) -- (0.5,0.5);

\draw[-,red] (5,5) -- (5.2,4.8);
\draw[-,red] (4.5,4.5) -- (4.7,4.3);
\draw[-,red] (4,4) -- (4.2,3.8);
\draw[-,red] (3.5,3.5) -- (3.7,3.3);
\draw[-,red] (3,3) -- (3.2,2.8);
\draw[-,red] (2.5,2.5) -- (2.7,2.3);
\draw[-,red] (2,2) -- (2.2,1.8);
\draw[-,red] (1.5,1.5) -- (1.7,1.3);
\draw[-,red] (1,1) -- (1.2,0.8);
\draw[-,red] (0.5,0.5) -- (0.7,0.3);

\draw[-,violet] (5,5) -- (4.8,5.2);
\draw[-,violet] (4.5,4.5) -- (4.3,4.7);
\draw[-,violet] (4,4) -- (3.8,4.2);
\draw[-,violet] (3.5,3.5) -- (3.3,3.7);
\draw[-,violet] (3,3) -- (2.8,3.2);
\draw[-,violet] (2.5,2.5) -- (2.3,2.7);
\draw[-,violet] (2,2) -- (1.8,2.2);
\draw[-,violet] (1.5,1.5) -- (1.3,1.7);
\draw[-,violet] (1,1) -- (0.8,1.2);
\draw[-,violet] (0.5,0.5) -- (0.3,0.7);

\draw[-] (5.5,0.5) -- (10.5,0.5);
\draw[-] (5.5,5.5) -- (10.5,0.5);
\draw[->,red,thick] (5.5,5.5) -- (8,0.5); 
\draw (6.6,2.6) node {$d_2$};
\draw[->] (6.65,3) -- (5.6,3); 
\draw[->] (6.85,3) -- (7.9,3);

\draw[-] (5.5,10.5) -- (10.5,5.5);
\draw[->,blue,thick] (5.5,5.5) -- (8,8); 
\draw (7.5,6.9) node {$d_2$};
\draw[->] (6.6,6.4) -- (7.4,5.6); 
\draw[->] (6.4,6.6) -- (5.6,7.4); 

\draw[-,blue] (5.5,6) -- (5.7,6.2);
\draw[-,blue] (5.5,6.5) -- (5.7,6.7);
\draw[-,blue] (5.5,7) -- (5.7,7.2);
\draw[-,blue] (5.5,7.5) -- (5.7,7.7);
\draw[-,blue] (5.5,8) -- (5.7,8.2);
\draw[-,blue] (5.5,8.5) -- (5.7,8.7);
\draw[-,blue] (5.5,9) -- (5.7,9.2);
\draw[-,blue] (5.5,9.5) -- (5.7,9.7);
\draw[-,blue] (5.5,10) -- (5.7,10.2);
\draw[-,blue] (5.5,10.5) -- (5.7,10.7);

\draw[-,blue] (5.5,5.5) -- (5.7,5.7);

\draw[-,blue] (6,5.5) -- (6.2,5.7);
\draw[-,blue] (6.5,5.5) -- (6.7,5.7);
\draw[-,blue] (7,5.5) -- (7.2,5.7);
\draw[-,blue] (7.5,5.5) -- (7.7,5.7);
\draw[-,blue] (8,5.5) -- (8.2,5.7);
\draw[-,blue] (8.5,5.5) -- (8.7,5.7);
\draw[-,blue] (9,5.5) -- (9.2,5.7);
\draw[-,blue] (9.5,5.5) -- (9.7,5.7);
\draw[-,blue] (10,5.5) -- (10.2,5.7);
\draw[-,blue] (10.5,5.5) -- (10.7,5.7);

\draw[-,violet] (5.5,6) -- (5.3,6.2);
\draw[-,violet] (5.5,6.5) -- (5.3,6.7);
\draw[-,violet] (5.5,7) -- (5.3,7.2);
\draw[-,violet] (5.5,7.5) -- (5.3,7.7);
\draw[-,violet] (5.5,8) -- (5.3,8.2);
\draw[-,violet] (5.5,8.5) -- (5.3,8.7);
\draw[-,violet] (5.5,9) -- (5.3,9.2);
\draw[-,violet] (5.5,9.5) -- (5.3,9.7);
\draw[-,violet] (5.5,10) -- (5.3,10.2);
\draw[-,violet] (5.5,10.5) -- (5.3,10.7);

\draw[-,violet] (5.5,5.5) -- (5.7,5.7);

\draw[-,red] (5.5,5.5) -- (5.7,5.3);

\draw[-,red] (6,5.5) -- (6.2,5.3);
\draw[-,red] (6.5,5.5) -- (6.7,5.3);
\draw[-,red] (7,5.5) -- (7.2,5.3);
\draw[-,red] (7.5,5.5) -- (7.7,5.3);
\draw[-,red] (8,5.5) -- (8.2,5.3);
\draw[-,red] (8.5,5.5) -- (8.7,5.3);
\draw[-,red] (9,5.5) -- (9.2,5.3);
\draw[-,red] (9.5,5.5) -- (9.7,5.3);
\draw[-,red] (10,5.5) -- (10.2,5.3);
\draw[-,red] (10.5,5.5) -- (10.7,5.3);

%\draw (5.5,10.5) node[left] {$1$};
%\draw (10.5,5.5) node[below] {$1$};
%\draw (5.5,0.5) node[left] {$-1$};
%\draw (0.5,5.5) node[below] {$-1$};

\end{tikzpicture}
\end{center}
\caption{\label{fig:classA3} Class $A_3$ with $
D(A_3)=\left(
\begin{array}{ccc}
a_1 & 1-a_2 & -1 \\
-1 & a_2 & 1-a_3 
\end{array} \right).
$}
\end{figure}

\paragraph{Case 3} All diagonal payoff matrix entries $a_{ii}=a_i$ for $i \in \{1,2,3\}$. Without loss of generality (considering strategy relabelling) we can assume that the matrix is of the following structure.

\[A=\left(
\begin{array}{ccc}
a_1 &  &  \\
0 & a_2 &  \\
1 &  & a_3
\end{array} \right)\] \medskip

\noindent Case 3.1: We fill the second and third columns as follows, which gives us
\[\left(
\begin{array}{ccc}
a_1 & 0 & 0  \\
0 & a_2 & 1 \\
1 & 1 & a_3
\end{array} \right)\] 
In this game strategy $1$ is dominated by strategy $3$. \medskip

\noindent Case 3.1: We fill the second and third columns as follows, which gives us
\[\left(
\begin{array}{ccc}
a_1 & 0 & 0  \\
0 & a_2 & 1 \\
1 & 1 & a_3
\end{array} \right)\] 
In this game strategy $1$ is dominated by strategy $3$ and there can be no completely mixed equilibrium. \medskip

\noindent Case 3.2: We fill the second and third columns as follows, which gives us
\[\left(
\begin{array}{ccc}
a_1 & 0 & 1  \\
0 & a_2 & 0 \\
1 & 1 & a_3
\end{array} \right)\] 
In this game strategy $2$ is dominated by strategy $3$ and there can be no completely mixed equilibrium. \medskip

\noindent Case 3.3: We fill the second and third columns as follows, which gives us $A_3$:
\[A_3=\left(
\begin{array}{ccc}
a_1 & 1 & 0  \\
0 & a_2 & 1 \\
1 & 0 & a_3
\end{array} \right)\] 
with
\[D(A_3)=\left(
\begin{array}{ccc}
a_1 & 1-a_2 & -1 \\
-1 & a_2 & 1-a_3 
\end{array} \right).\]
This game has no dominated strategies. Figure \ref{fig:classA3} provides a geometric representation of $D(A_3)$, which is very similar to the one for $D(A_1)$ in Figure \ref{fig:classA1}. As in that case, the union of half-spaces clearly covers all of $\R^2$ for all parameter configurations. By Corollary \ref{cor_nec_cond_completelymixed_Nash_eq} this implies that the game has completely mixed symmetric equilibrium. To see that it is the only symmetric equilibrium, note that the game has no symmetric pure strategy equilibria and for any mixed strategy that uses only two pure strategies, one of these two pure strategies then strictly dominates the other one (in the game with the unused strategy removed).\medskip

\noindent Case 3.4: We fill the second and third columns as follows, which gives us
\[\left(
\begin{array}{ccc}
a_1 & 1 & 1  \\
0 & a_2 & 0 \\
1 & 0 & a_3
\end{array} \right)\] 
In this game strategy $2$ is dominated by strategy $1$ and there can be no completely mixed equilibrium. \medskip

\section{Related Literature}

Farkas' lemma has played an important role in several areas of economic theory, including game theory, through the relationship between equilibria (in games) and solutions to linear programming problems, see e.g. \cite{brooks2023canonical}. We here use Farkas' lemma for identifying a convenient necessary and sufficient condition for a player in a two-player game to be completely indifferent between all strategies. This condition is in terms of properties of the column vectors of the matrix of this player's payoff differences between consecutive pure strategies. This condition, in turn, delivers a necessary condition for the existence of a completely mixed Nash equilibrium in such games. 

There are several papers identifying necessary and/or sufficient conditions for completely mixed equilibria.  \cite{kaplansky1945contribution} and \cite{kaplansky1995contribution} identify necessary and sufficient conditions for zero-sum games to have a completely mixed Nash equilibrium in terms of cofactors of the payoff matrix. \cite{parthasarathy2020completely} extends these results to $3 \times 3$ non-zero sum bimatrix games with skew symmetric payoff matrices. \cite{raghavan1970completely} provides a necessary condition in terms of the rank of the payoff matrix, but the condition is not sufficient. \cite{milchtaich2006computation} and \cite{milchtaich2008some} compute expected payoffs in completely mixed equilibria and provide a necessary and sufficient condition for a unique completely-mixed Nash-equilibrium in terms of the determinants of certain transformations of the payoff matrices. \citet[Corollary 1]{weinstein2020best} shows that a two-player game has a totally mixed Nash equilibrium if and only if neither player has a pair of mixed strategies such that one weakly dominates the other. 

All these conditions for the existence of completely mixed equilibria are in terms of different properties of the game. Any one of these has their uses. Sometimes one will be useful, sometimes another. We found our necessary condition in Corollary~\ref{cor_nec_cond_completelymixed_Nash_eq} helpful to prove some results in a recent paper of ours (\cite{herold2020evolution}) and we hope it turns out to be useful for other game theorists as well.

\bibliographystyle{chicago}
\bibliography{evolution.bib}

%\printbibliography

\end{document}